\documentclass[prd,superscriptaddress,twocolumn,floatfix,footinbib]{revtex4-1}

\usepackage[plainpages=false,pdfpagelabels,colorlinks=true,linkcolor=red,urlcolor=blue,citecolor=blue,
pdftitle={Typical entropy of a subsystem: Page curve and its variance},
pdfauthor={Eugenio Bianchi},pdfdisplaydoctitle=true,pdfduplex=DuplexFlipLongEdge]{hyperref}

\usepackage{amsmath,amsfonts,amssymb,amsthm}
\usepackage{graphicx}
\usepackage{hyperref}
\usepackage{dsfont}
\usepackage{color}

\newcommand{\ket}[1]{|#1\rangle}
\newcommand{\bra}[1]{\langle #1|}
\newcommand{\av}[1]{\langle #1\rangle}
\newcommand{\Tr}{\mathrm{Tr}}
\renewcommand{\dim}{\mathrm{dim}}

\newcommand{\dd}{\mathrm{d}}

\renewcommand{\H}{\mathcal{H}}
\renewcommand{\l}{\lambda}
\newcommand{\Smax}{S_{\mathrm{max}}}

\newcommand{\mm}{m} 

\newcommand{\PRLsep}{\noindent\makebox[\linewidth]{\resizebox{0.3333\linewidth}{1pt}{$\bullet$}}\medskip}

\begin{document}

\title{Typical entanglement entropy in the presence of a center: Page curve and its variance}

\author{Eugenio Bianchi}
\affiliation{Institute for Gravitation and the Cosmos, The Pennsylvania State University, University Park, Pennsylvania 16802, USA}
\affiliation{Department of Physics, The Pennsylvania State University, University Park, Pennsylvania 16802, USA}

\author{Pietro Don\`a}
\affiliation{Institute for Gravitation and the Cosmos, The Pennsylvania State University, University Park, Pennsylvania 16802, USA}
\affiliation{Department of Physics, The Pennsylvania State University, University Park, Pennsylvania 16802, USA}

 \begin{abstract}
In a quantum system in a pure state, a subsystem generally has a nonzero entropy because of entanglement with the rest of the system. Is the average entanglement entropy of pure states also the typical entropy of the subsystem? We present a method to compute the exact formula of the momenta of the probability $P(S_A) \mathrm{d}S_A$ that a subsystem has entanglement entropy $S_A$. The method applies to subsystems defined by a subalgebra of observables with a center. In the case of a trivial center, we reobtain the well-known result for the average entropy and the formula for the variance. In the presence of a nontrivial center, the Hilbert space does not have a tensor product structure and the well-known formula does not apply. We present the exact formula for the average entanglement entropy and its variance in the presence of a center. We show that for large systems the variance is small, $\Delta S_A/\langle{S_{A}}\rangle\ll 1$, and therefore the average entanglement entropy is typical. We compare exact and numerical results for the probability distribution and comment on the relation to previous results on concentration of measure bounds. We discuss the application to physical systems where a center arises. In particular, for a system of noninteracting spins in a magnetic field and for a free quantum field, we show how the thermal entropy arises as the typical entanglement entropy of energy eigenstates.
 \end{abstract}

\maketitle

\setlength{\abovedisplayskip}{.7em}
\setlength{\belowdisplayskip}{.7em}

\begin{figure}[t]
\centering
\includegraphics[width=0.47\textwidth]{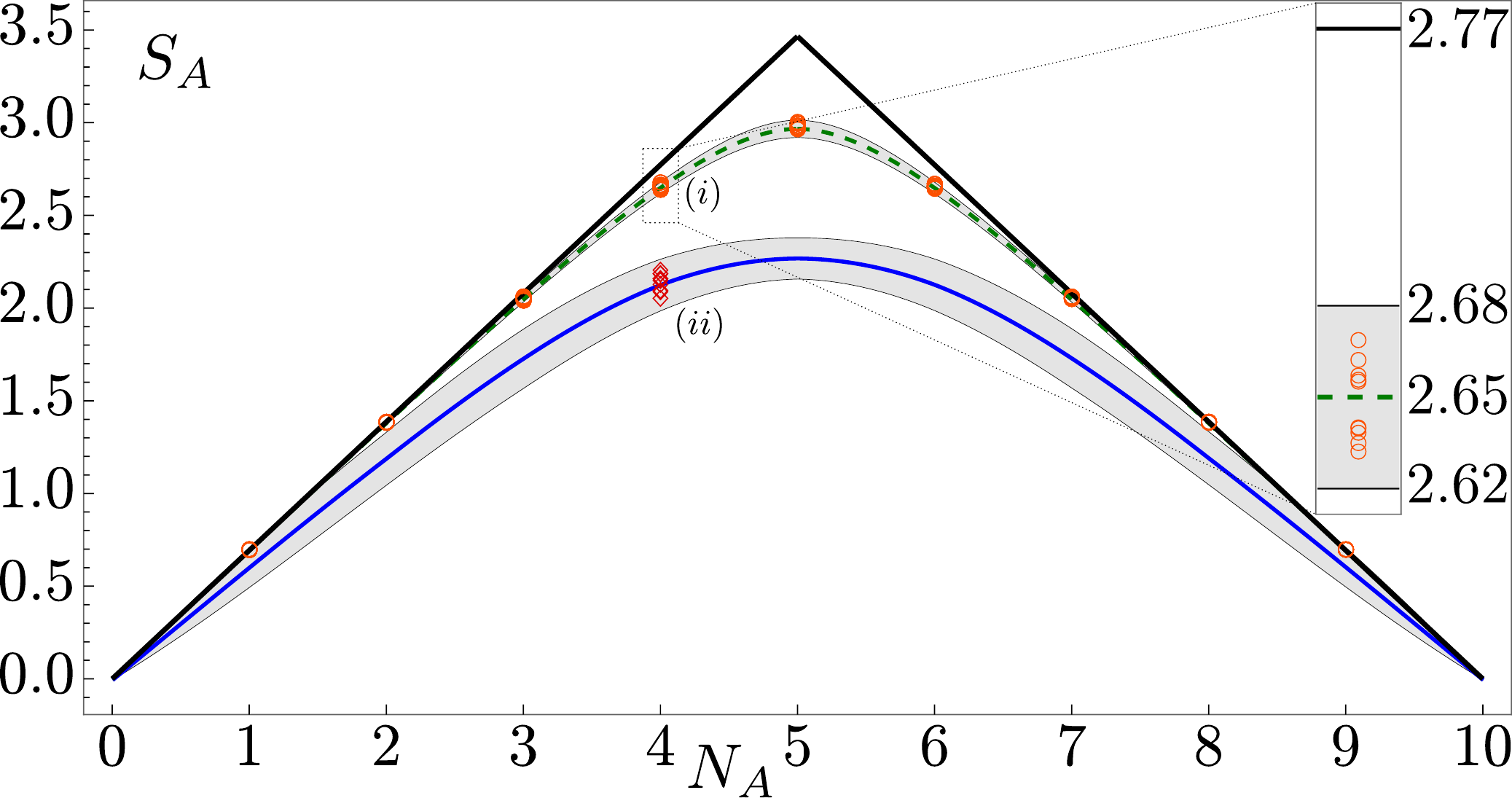}
\caption{\emph{Page curve} of a system of  $N=10$ noninteracting spins. (i) \emph{Trivial center}. In the absence of an external magnetic field, all states are equiprobable: the entanglement entropy $S_A$ of a sample of random pure states is shown as a function of the subsystem size $N_A$ (orange circles), together with the band $\av{S_A}\pm 3 \,\Delta S_A$
(green dashed line and gray region). The inset shows the data for $N_A=4$. (ii) \emph{nontrivial center}. In the presence of a magnetic field, the entanglement entropy of random pure states of given energy $E_n=\mu_0 \mathrm{B}\,n$ with $n=-2$ is shown for $N_A=4$ (red diamonds), together with the band $\av{S_A}_{\!{}_E}\pm \,3\,(\Delta S_A)_{\!{}_E}$ for the Page curve at fixed energy  (blue solid line and gray region).}
\label{fig:pagecurve} 
\end{figure}

\emph{Introduction.}---In a seminal paper \cite{Page:1993df}, Page showed that, when an isolated quantum system is in a random pure state, the average entropy of a subsystem is close to maximal. This result plays a central role in the analysis of the black hole information puzzle \cite{Page:1993wv,Giddings:2012bm,Braunstein:2009my,Almheiri:2012rt,Marolf:2017jkr,Harlow:2014yka,Bianchi:2014bma,Abdolrahimi:2015tha}, in the quantum foundations of statistical mechanics
\cite{Deutsch1991, Srednicki1994, rigol.2008,d2016quantum, gogolin2016equilibration,Deutsch2018,goldstein_lebowitz_06, popescu_short_06, tasaki_98,polkovnikov2011colloquium,Vidmar:2017uux,Vidmar:2018rqk,Hackl:2018tyl,Vidmar:2017abc}, in quantum information theory \cite{Hayden:2006,Hayden:2007cs,Sekino:2008he,Hosur:2015ylk,Roberts:2016hpo,Fujita:2017pju,Lu:2017tbo,Fujita:2018wtr} and in the study of the quantum nature of spacetime geometry \cite{VanRaamsdonk:2010pw,Bianchi:2012ev,Jacobson:2015hqa,Bianchi:2018fmq,Baytas:2018wjd,Qi:2018ogs}. The entanglement entropy, expressed as a function of the subsystem size, is often called the \emph{Page curve} (Fig.~\ref{fig:pagecurve}).

In a quantum system in a pure state, a subsystem $A$ generally has a nonzero entropy because of entanglement with the rest of the system. In this paper we address the question: \emph{Is the average entanglement entropy of pure states $\av{S_A}$ also the typical entropy of the subsystem?}
To illustrate the significance of this question, let us consider for instance the gas in a room held at fixed temperature. The canonical ensemble allows us to compute the average energy of the gas. However, the configuration of molecules in one room is one realization of this ensemble---we are not averaging over rooms. How close to the average is the energy of this realization? In other words, is the average energy typical? In statistical mechanics, we answer this question by computing the variance $(\Delta E)^2$. In the canonical ensemble, we find $\Delta E/\hspace{-1pt}\av{E}\ll 1$ and we conclude that the average energy is typical. Here we investigate the typicality of the entanglement entropy $S_A$ of a subsystem by studying its average and variance.

A well-studied special case of subsystem $A$ is described by a subalgebra with a trivial center. In this case the Hilbert space of the system is simply given by a tensor product of the subsystem Hilbert space $\H_A$ and its complement $\H_B$, i.e. $\H = \H_{A} \otimes \H_{B}$. This case was originally considered by Page \cite{Page:1993df,Lubkin:1993,Lloyd:1988cn,Foong:1994,Sanchez:1995,Sen:1996ph,Dyer:2014} who computed the average in $\H$ of the entanglement entropy. The variance was computed recently in \cite{Vivo,Wei}. In this paper we present an algorithm that reproduces the average and variance, together with the skewness and higher moments of  the entanglement entropy probability distribution $P(S_A) \mathrm{d}S_A$ for a trivial center.

In many physically relevant cases, the Hilbert space over which we average the entanglement entropy does not have a tensor product structure, $\H \neq \H_{A} \otimes \H_{B}$. In fact, in general, the subalgebra that identifies the subsystem $A$ has a nontrivial center. As a result, the Hilbert space of the system is a direct sum of tensor products \cite{Casini:2013rba,Ma:2015xes}. In this work we extend the typicality results on the entanglement entropy in the presence of a center.  Examples of subsystems with a nontrivial center are subsets of a spin chain with fixed boundary conditions; lattice fermions with fixed boundary conditions \cite{Lin:2018bud}; compact scalar fields, Abelian and non-Abelian gauge fields on a lattice \cite{Casini:2013rba}; loop quantum gravity \cite{Rovelli:2004tv,Donnelly:2008vx,Bianchi:2018fmq,Baytas:2018wjd}.

A simple example of subsystems with a center is provided by the energy eigenspace in a noninteracting system with Hamiltonian $H=H_A+H_B$. In this case the eigenspace $\H(E)\subset \H$ has the structure of a direct sum of tensor products
\begin{equation}
\textstyle \H(E)=\bigoplus_{j=1}^J\Big(\H_A(\varepsilon_j)\otimes\H_B(E-\varepsilon_j)\Big)\,,
\label{eq:Hoplus}
\end{equation}
where $\H_A(\varepsilon_j)$ and $\H_B(\varepsilon_k)$ are eigenspaces of given energy for the subsystems $A$ and $B$. This structure is relevant for the study of  thermal properties of isolated quantum systems. See Fig.~\ref{fig:pagecurve} for an example.

Building on \cite{Page:1993df,Lubkin:1993,Lloyd:1988cn,Foong:1994,Sanchez:1995,Sen:1996ph,Dyer:2014}, we develop methods to compute the exact formula for the average entanglement entropy and the moments $\mm_n=\langle(S_A-\av{S_A})^n\rangle$ of the probability distribution $P(S_A) \dd S_A$ of finding a pure state with entanglement entropy $S_A$. Our methods are tailored to the computation of averages over pure states in $\H$ and over pure states in the eigenspace $\H(E)$.

The exact formula for the average $\av{S_A}$ over all pure states in $\H$ was conjectured by Page in \cite{Page:1993df}, improving earlier works \cite{Lubkin:1993,Lloyd:1988cn}, and later proved with different methods in \cite{Foong:1994,Sanchez:1995,Sen:1996ph,Dyer:2014} and in \eqref{eq:averageentropy}. Here we present the exact formula for the average $\av{S_A}_{\!{}_E}$ over all states in the Hilbert $\H(E)$ with a nontrivial center. Furthermore, we determine the exact formulas for: the variance $\mm_2$ for pure states in $\H$ reproducing the result found in \cite{Vivo,Wei} for a trivial center; the exact formula for third order moment $\mm_3$ with a trivial center; the exact formula for $\mm_2$ in the presence of a nontrivial center. We compare our results to concentration of measure bounds \cite{Hayden:2006}.

By studying the average entanglement entropy $\av{S_A}$ and the variance
\begin{equation}
(\Delta S_A)^2 = \av{{S_A}^2} - \av{S_A}{}^2,
\label{eq:vardef}
\end{equation}
we show that, both in the case of a trivial and a nontrivial center, for large systems the average entanglement entropy of a subsystem is also typical.

After presenting a detailed derivation of the Page curve and its variance, we discuss the application of our results to a model system where a nontrivial center arises: we determine the Page curve of energy eigenstates of a paramagnetic solid in a magnetic field (Fig.~\ref{fig:pagecurve}), and show how thermal properties of a subsystem arise from entanglement with the rest of the system.

\medskip

\emph{Average entropy and variance with a trivial center.}--- Let us consider a quantum system with a bipartite Hilbert space $\mathcal{H}=\H_A\otimes \H_B$. The subsystems $A$ and $B$ have dimension $d_A=\dim \H_A$ and $d_B=\dim \H_B$, with $A$ the smaller of the two subsystems, $1<d_A\leq d_B$, and the dimension of $\H$ given by  $d=d_A d_B$. The restriction of a pure state $\ket{\psi}\in \H$ to the subsystem $A$ defines the reduced density matrix $\rho_A=\Tr_{\!B}\ket{\psi}\bra{\psi}$. The entanglement entropy $S_{A}(\psi)=-\Tr(\rho_A\log \rho_A)$ is the von Neumann entropy of the reduced density matrix. 

A random state $|\psi\rangle$ in the Hilbert space $\H$ can be generated by choosing an orthonormal basis  $|n\rangle$ with $n=1,\ldots,d$ and picking a vector $|\psi\rangle=\sum_n \psi_n |n\rangle$ at random with respect to the uniform measure $\dd\mu(\psi)=Z^{-1}\delta(|\psi|^2-1)\dd \psi\dd\overline{\psi}$ on the unit sphere in $\mathbb{C}^d$, with the constant $Z$ defined so that the measure is normalized to unity, $\int \dd\mu(\psi) =1$. The average of a function $f$ over all states $\ket{\psi}$ can be computed by integrating uniformly over states, $\av{f}\,=\int f(\psi)\, \dd \mu(\psi)$. As the entanglement entropy $S_{A}$ depends on the reduced density matrix $\rho_{A}$ only via its eigenvalues $\lambda_a$, (the entanglement spectrum, with $a=1,\ldots, d_{A}$), to compute the average over $\ket\psi$ we need only  the induced measure over the eigenvalues $\lambda_a$. This measure $\dd \mu(\l_{1},\ldots,\l_{d_{A}})$ was computed by Lloyd and Pagels in \cite{Lloyd:1988cn}, (see also App.~\ref{App:A}).

Computing the average entanglement entropy $\langle S_A\rangle$ and its variance $\Delta S_A$ is not immediate because they are not polynomial functions of $\rho_A$. Here we follow a strategy that generalizes to the computation of higher moments of $S_A$. We first consider the average of the function $\av{\Tr_{\!A}(\rho_{A}^{r})}$ with $r\geq 0$,
\begin{equation}
\label{eq:tracerhor}
\av{\Tr(\rho_{A}^{r})}=\int\textstyle\big(\sum_{a=1}^{d_{A}}\l_{a}^{r}\big) \,\dd \mu(\l_{1},\ldots,\l_{d_{A}}) \, .
\end{equation}
To compute the integral over $\lambda_a$ it is useful to introduce the quantity $X_{ij}(r)$,  defined as an integral of generalized Laguerre polynomials $L_{i}^{(d_{B}-d_{A})}(q)$ with  $i=0,\ldots,d_A-1$:
\begin{align}
&X_{ij}(r)\, =\frac{\int_{0}^{\infty}q^{r} L_{i}\!\!{}^{{}^{(d_{B}-d_{A})}}\!(q)\,L_{j}\!\!{}^{{}^{(d_{B}-d_{A})}}\!(q)\,q^{d_{B}-d_{A}}e^{-q}\,\dd q}{\Gamma(i+1)\Gamma(d_B-d_A+i+1)}\nonumber 
 \\[.5em]
& =\sum_{p=0}^{d_A-1}\textstyle\frac{\Gamma(d_B-d_A+r+1+p)\,\Gamma(j+1)\Gamma(r+1)^{2}/\Gamma(d_B-d_A+i+1)}{\Gamma(i-p+1) \Gamma(r+p-i+1)\Gamma(j-p+1)\Gamma(r-j+p+1)\Gamma(p+1)} 
\label{eq:matrixX}
\end{align}
Computing the integral \eqref{eq:matrixX} is nontrivial. We obtained the result by using the generating function for the Laguerre polynomials, as we describe in more detail in App.~\ref{App:A}. Formula \eqref{eq:matrixX} for $X_{ij}(r)$ provides the main technical tool for our derivation.

The average \eqref{eq:tracerhor} takes a simple form when expressed in terms of the matrix $X_{ij}(r)$,
\begin{equation}
\label{eq:rho1}
\av{\Tr(\rho_{A}^{r})}= \frac{\Gamma(d_{A}d_{B})}{\Gamma(d_{A}d_{B}+r)}\Tr X(r) \,,
\end{equation}
where we treat $X_{ij}(r)$ as a $d_A\times d_A$ matrix $X(r)$. We note that \eqref{eq:rho1} is a smooth continuous function of $r$. The average entanglement entropy can be obtained immediately by taking a derivative with respect to $r$,
\begin{equation}
\label{eq:trick1}
\av{S_{A}}=-\av{\Tr\,\rho_{A}\log\rho_{A}}= - \lim_{r\to1}\partial_r \av{\Tr(\rho_{A}^{r})} \, .
\end{equation}
The result of this computation, expressed in terms of the digamma function $\Psi(x)=\Gamma'(x)/\Gamma(x)$ (i.e., the logarithmic derivative of the gamma function), is the formula
\begin{equation}
\label{eq:averageentropy}
\av{S_{A}}= \Psi(d_{A}d_{B}+1)-\Psi(d_{B}+1)-\frac{d_{A}-1}{2d_{B}} \,.
\end{equation}
This formula was first conjectured by Page \cite{Page:1993df} and then proved with different techniques here and in \cite{Foong:1994,Sanchez:1995,Sen:1996ph,Dyer:2014}.

The technique described above applies directly to the calculation of the average of $S_A{}^2$. The strategy is to first consider the average of $\av{\Tr(\rho_{A}^{r_1})\,\Tr(\rho_{A}^{r_2})}$ and express it in terms of matrices $X(r)$:
\begin{align}
& \av{\Tr(\rho_{A}^{r_1})\,\Tr(\rho_{A}^{r_2})}=
\frac{\Gamma(d_A d_B)}{\Gamma(d_A d_B+r_{1}+r_{2})}\times\label{eq:rho2}\\
&\!\times\!\big(\Tr X(r_1\!+\!r_{2})+\Tr X(r_{1})\,\Tr X(r_{2})-\Tr\big( X(r_{1})X(r_{2})\big)\big).\nonumber
\end{align}
Its derivatives with respect to $r_1$ and $r_2$ can again be expressed in terms of derivatives of the gamma function, and simplified with the help of Wolfram's Mathematica. The average of $S_A{}^2$ is obtained as
\begin{equation}
\av{S_A{}^2}=\!\lim_{\quad r_1,r_2\to 1}\partial_{r_1}\partial_{r_2}\av{\Tr(\rho_{A}^{r_{1}})\Tr(\rho_{A}^{r_{2}})} \, .
\end{equation}
Using the definition \eqref{eq:vardef} and the formula \eqref{eq:averageentropy}, we obtain the exact formula:
\begin{align}
&\textstyle (\Delta S_A)^2=\;\frac{d_{A}+d_{B}}{d_{A}d_{B}+1}\Psi'(d_{B}+1)\,+ \label{eq:variance}\\[.5em]
&\textstyle \qquad\qquad-\Psi'(d_{A}d_{B}+1)-\frac{(d_{A}-1)(d_{A}+2d_{B}-1)}{4d_{B}^{2}(d_{A}d_{B}+1)}\,,\nonumber
\end{align}
discussed also in \cite{Vivo,Wei}. Using the same technique we can determine also the average of higher powers of $S_A$. In particular, we report the exact formula for the third moment $\mm_3$ of the entanglement entropy distribution (App.~\ref{App:A}) 
\begin{align}
& \mm_3\;=\;\textstyle \Psi''(d_{A}d_{B}+1)-\frac{d_{A}^{2}+3d_{A}d_{B}+d_{B}^{2}+1}{(d_{A}d_{B}+1)(d_{A}d_{B}+2)}\Psi''(d_{B}+1)+\nonumber \\[.5em]
 &\qquad \textstyle +\frac{(d_{A}^{2}-1)(d_{A}d_{B}-3d_{B}^{2}+1)}{d_{B}(d_{A}d_{B}+1)^{2}(d_{A}d_{B}+2)}\Psi'(d_{B}+1)+\nonumber\\[.5em]
&\qquad \textstyle -\frac{(d_{A}-1)(2d_{A}^{3}d_{B}+3d_{A}^{2}d_{B}^{2}-4d_{A}^{2}d_{B}+4d_{A}d_{B}^{3}-3d_{A}d_{B}^{2})}{4d_{B}^{3}(d_{A}d_{B}+1)^{2}(d_{A}d_{B}+2)}+\nonumber\\[.5em]
&\qquad \textstyle -\frac{(d_{A}-1)(2d_{A}^{2}+8d_{A}d_{B}+10d_{B}^{2}-4d_{A}-6d_{B}+2)}{4d_{B}^{3}(d_{A}d_{B}+1)^{2}(d_{A}d_{B}+2)}\,.\label{eq:skew}
\end{align}
We observe that the formulas for the moments $\mm_n$ are exact: they provide the mean \eqref{eq:averageentropy}, the variance \eqref{eq:variance} and the skewness \eqref{eq:skew} of the entropy for small systems, as well as for large systems. In the latter case, a Taylor series  in $1/d_B$ provides asymptotic expressions for large systems and any subsystem: the average entropy is approximated by the expression
\begin{equation}
\av{S_{A}}\approx \log d_A - \frac{1}{ d_A d_B}\,\frac{d_A^{2}-1}{2}\quad\; \textrm{for}\quad d_B \gg 1\,.
\label{eq:average_ldb}
\end{equation}
As the entropy of $A$ can be at most $\Smax=\log d_{A}$, this formula shows that for a large system the average entropy of a subsystem is close to maximal. The asymptotic expression for the variance of $S_A$ is:
\begin{equation}
\Delta S_A\;\approx\; \frac{1}{ d_A d_B}\sqrt{\frac{d_A^{2}-1}{2}} \qquad \textrm{for}\qquad d_B \gg 1\,.
\label{eq:variance_ldb}
\end{equation}
We observe that the variance vanishes in the limit of $d_B\to \infty$. Therefore we conclude that, in a large system, any subsystem has $\Delta S_A/\av{S_A}\ll 1$. In other words, in a large system, the average entropy of any subsystem is also its typical entropy. We report also the asymptotic expression for the third moment $\mm_3$ \eqref{eq:skew}
\begin{equation}
\mm_3\approx -\frac{d_A^2-1}{d_A^3 d_B^3} \, \qquad\textrm{for}\quad d_B\gg 1\,.
\label{eq:mu3}
\end{equation}
This formula shows that the skewness $\mm_3/\sigma^3\approx  -\sqrt{8/(d_A^2-1)}$ is negative, which results in a right tilt in the distribution that does not vanish as $d_B\to \infty$.

One might expect that these asymptotic formulas can be obtained in a simpler way, for instance by first expanding the entropy $S_A$ around the maximally mixed state and then taking the average. The expansion in $\delta \rho_A= \rho_A-\mathds{1}/d_A$ was first proposed by Lubkin and Lubkin in \cite{Lubkin:1993} and is commonly found in reviews \cite{Harlow:2014yka}. However, it was shown by Dyer in \cite{Dyer:2014} that the series in $\delta \rho_A$ for $\av{S_A}$ does not converge. This has the consequence that, truncating the expansion in  $\delta \rho_A$, only the first order in Page's formula \eqref{eq:average_ldb} is accidentally reproduced. Similarly, the leading order variance $\Delta S_A$ \eqref{eq:variance_ldb} cannot be obtained by truncating the expansion in $\delta \rho_A$.

\begin{figure*}[t!]
\includegraphics[width=0.97\textwidth]{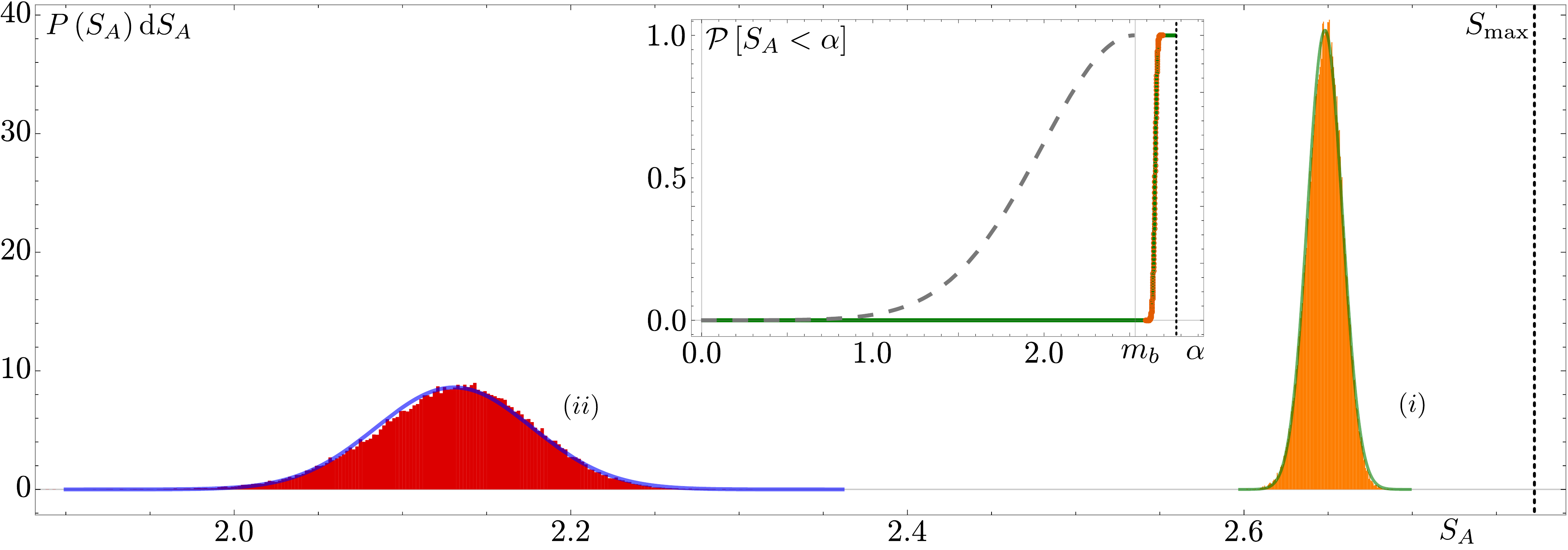}
\caption{\label{fig:Histogram} (i) \emph{Trivial center}. Probability of finding entanglement entropy $S_A$ in a sample of $10^5$ random pure states of a spin system ($N=10$ and $N_A=4$, in orange 200 bins). See Fig.~\ref{fig:pagecurve} (inset) for reference. We compare the sample to the normal probability distribution \eqref{eq:normalprobability} with mean \eqref{eq:averageentropy} and variance \eqref{eq:variance} (solid green line). \emph{Inset: Cumulative probability distribution} of finding entanglement entropy smaller than $\alpha$ for the same sample (orange circles). We compare the sample to the normal cumulative distribution (solid green line) and to the concentration of measure bound \eqref{eq:Hayden} (dashed gray line). (ii) \emph{nontrivial center}. Probability of finding entanglement entropy $S_A$ in a sample of $10^5$ random pure states of energy $E_n=\mu_0 \mathrm{B}\, n$ with $n=-2$ of a spin system in a magnetic field ($N=10$ and $N_A=4$, in red 200 bins). See Fig.~\ref{fig:pagecurve} (red diamonds) for reference. We compare the sample to the normal probability distribution with mean \eqref{eq:SavOplus} and variance \eqref{eq:variance} (solid blue line)}
\end{figure*}

Determining the full probability distribution $P(S_A) \dd S_A$ of a random pure state is not immediate. The methods introduced here allow us to determine its average $\mm = \av{S_A}$, its variance $\sigma^2 = (\Delta S_A)^2$ and higher order moments $\mm_n$ such as the skewness \eqref{eq:mu3}, (see also App.~\ref{App:A}). The normal distribution 
\begin{equation}
\textstyle P_{\mathcal{N}}(S_A)\, \dd S_A = \frac{1}{\sqrt{2\pi} \sigma} \exp\big({-\frac{\left(S_A - \mm \right)^2}{2 \sigma^2}}\big) \dd S_A
\label{eq:normalprobability}
\end{equation}
is the distribution with the largest Shannon entropy at fixed average $\mm$ and variance $\sigma^2$. In Fig.~\ref{fig:Histogram} we compare this distribution to a numerical sample and find that it characterizes well the support of the probability distribution $P(S_A) \dd S_A$ of random pure states. 
In particular, the numerical sample and the analytic formulas \eqref{eq:average_ldb}, \eqref{eq:variance_ldb} show that it is unlikely to find a state with maximum entropy $\Smax$. 
The numerical sample shows also a small right tilt with respect to the normal distribution, in accordance with the negative skewness \eqref{eq:mu3}.

Previous analysis of typicality have used concentration of measure techniques to  provide upper bounds on the probability of finding a state with entropy lower than the average entropy \cite{Hayden:2006}. In particular, using Levy's lemma,  Theorem III.3 in \cite{Hayden:2006}\cite{notapp} states that the cumulative distribution $\mathcal{P}[ S_A < \alpha\,]$ is bounded from above by a Gaussian function, 
\begin{equation}
 \mathcal{P}[ S_A < \alpha\,]=\! \int_0^\alpha\!\! P( S_A ) \dd S_A\, \leq\, \textstyle\exp\! \Big(\!\!-\frac{(\alpha - \mm_b)^2}{2\sigma_b^2} \Big) 
\label{eq:Hayden}
\end{equation}
with $\mm_b = \log d_A - \frac{d_A}{d_B}$, $\sigma_b = \frac{2 \pi \log d_A}{\sqrt{d_A d_B -1}}$ and $\alpha<\mm_b$. 
We compare this bound to the normal cumulative distribution, and to a numerical sample of random pure states. The inset in Fig.~\ref{fig:Histogram} clearly shows  that the probability is more concentrated than what the bound \eqref{eq:Hayden} indicates.

\bigskip

\emph{Average entropy and variance in the presence of a \mbox{nontrivial} center}.--- Consider a system with algebra of observables $\mathcal{A}$. We define a subsystem $A$ by choosing a subalgebra of observables $\mathcal{A}_A$. The complement of the subsystem $A$ is denoted $B$ and its algebra of observables is the commutant of $\mathcal{A}_A$, i.e., $\mathcal{A}_B=\{b \in \mathcal{A} \ |  \ [b,a] =0 \ \forall a \in \mathcal{A}_A\}$. The intersection of the two subalgebras, $\mathcal{A}_A \cap \mathcal{A}_B=\mathcal{Z}_A$, is called center of $\mathcal{A}_A\subset \mathcal{A}$. In the presence of a center, the Hilbert space of the system decomposes as a direct sum of tensor products \cite{Casini:2013rba,Ma:2015xes},
\begin{equation}
\textstyle \H = \bigoplus_{\zeta} \Big(\H_A(\zeta)\otimes\H_B(\zeta)\Big)\,,
\label{eq:Hoplus2}
\end{equation}
where the sum is over the spectrum of $\mathcal{Z}_A$.

We give a concrete example of a center. Let us consider a composite system with Hilbert space $\H=\H_A\otimes \H_B$ and Hamiltonian  $H=H_A+H_B$ having energy eigenvalues $E_{jk}=\varepsilon_{jA}+\varepsilon_{kB}$. On the energy eigenspace $\H(E)\subset\H$ the algebra of the subsystem $A$ has a nontrivial center $\mathcal{Z}_A = H_A$. Therefore, $\H(E)$ has the structure \eqref{eq:Hoplus2},
\begin{equation}
\textstyle \H(E)=\bigoplus_{j=1}^J\Big(\H_A(\varepsilon_j)\otimes\H_B(E-\varepsilon_j)\Big)\,,
\label{eq:Hoplus4}
\end{equation}
where the sum over $j$ is such that $\varepsilon_{kB}=E-\varepsilon_{jA}$.

Energy eigenspaces of the subsystem $A$ are denoted $\H_{A}(\varepsilon_j)$ and have dimension $d_{jA}=\dim \H_{A}(\varepsilon_j)$. Similarly for subsystem $B$. The energy eigenspaces of the system have then the direct sum structure $\H(E)=\oplus_j \H_j(E)$ where the sector $\H_j(E)=\H_A(\varepsilon_j)\otimes \H_B(E-\varepsilon_j)$ has definite energy in each subsystem. We denote $d_j=\dim \H_j(E)$ the dimension of each sector, with $d_j=d_{jA}\, d_{jB}$ and $d_E=\sum_j d_j$ the dimension of $\H(E)$.

Due to the direct-sum structure  \eqref{eq:Hoplus4}, Page's formula \eqref{eq:averageentropy} does not apply. Other instances of systems where the relevant Hilbert space has the form \eqref{eq:Hoplus}  are subsystems in lattice gauge theory \cite{Casini:2013rba}, in spin chains and in lattice fermions with fixed boundary conditions \cite{Lin:2018bud}, in loop quantum gravity \cite{Rovelli:2004tv,Donnelly:2008vx,Bianchi:2018fmq,Baytas:2018wjd}, and in general in presence of an additive constraint. The formulae that we derive below apply equally to all these cases.

To investigate typicality of the entropy in the energy eigenspace $\H(E)$, we determine the uniform measure over pure states belonging it. We note that a state $\ket{\psi,E}$ in $\H(E)$ can be written as a superposition $\ket{\psi,E}=\sum_j \sqrt{p_j}\,\ket{\phi_j}$ of normalized states $\ket{\phi_j}\in \H_j(E)$, with weights $p_j\geq 0$ satisfying the normalization condition $\sum_j p_j=1$. The reduced density matrix for the subsystem $A$ is given by $\rho_A=\sum_j p_j \rho_{jA}$ with $\rho_{jA}=\Tr_{B}\ket{\phi_j}\bra{\phi_j}$. The entanglement entropy of the subsystem, $S_A(\psi)=-\Tr(\rho_A\log \rho_A)$, splits into the sum of two terms \cite{Ohya2004}:
\begin{equation}
\textstyle S_A(\psi)=\;\sum_j p_j \,S_{jA}(\phi_j)\;-\sum_j p_j\log p_j\,,
\end{equation}
where $S_{jA}(\phi_j)=-\Tr(\rho_{jA}\log \rho_{jA})$ is the entanglement entropy in the sector $\H_j(E)$. We note that the entanglement entropy of $A$ is the $p_j$-weighted sum of the entanglement entropy $S_{jA}$ in each sector, plus the Shannon entropy of the weights $p_j$.

The uniform measure $\dd\mu_E(\psi)$ over pure states  in $\H(E)$ decomposes as  
\begin{equation}
\textstyle \dd\mu_E(\psi)= \dd\nu(p_1,\ldots,p_J)  \prod_j\dd \mu(\phi_j)\,,
\end{equation}
where $\dd \mu(\phi_j)$ is the uniform measure over pure states in each sector $\H_j(E)$. We derive the measure over the weights $p_j$ in App.~\ref{App:B} and find
\begin{equation}
\dd\nu(p_1,\ldots,p_J)=\frac{1}{\mathcal{Z}}\delta\big(\textstyle\sum_j p_j - 1 \big) \prod_j\!\big( p_j^{d_j-1}\dd p_j  \big)\,.
\end{equation}
This measure defines the Dirichlet distribution, also known as the multivariate beta distribution \cite{Balakrishnan}. The constant $\mathcal{Z}$ is defined so that the measure is normalized to unity, $\int \dd\nu(p_1,\ldots,p_J)=1$. The average weight is $\av{p_j}= d_j/d_E$ and its second moments are $\av{p_i\,p_j}=(d_i d_j+\delta_{ij}d_i)/(d_E(d_E+1))$.

Using the technique described in \eqref{eq:trick1}, we write the average in $\H(E)$ of the entanglement entropy as
\begin{equation}
\av{S_A}_{\!{}_E}=- \lim_{r\to1}\textstyle \partial_r\av{\sum_j p_j\Tr(\rho_{jA}^{\,r})+\sum_jp_j^r} \,,
\end{equation}
and find the exact formula
\begin{equation}
\textstyle \av{S_A}_{\!{}_E}=
\sum_j \!\frac{d_j}{d_{E}} \big(\av{S_{jA}}+\Psi(d_E+1)  -\Psi(d_j +1)\big)\,,
\label{eq:SavOplus}
\end{equation}
where $\av{S_{jA}}$ is given by \eqref{eq:averageentropy}. Using the same technique, we compute the exact formula for the average of $S_{A}{}^{2}$,
\begin{widetext}
${}$\\[-2.5em]
\begin{align}
\av{S_{A}{}^{2}}_{\!{}_E} =&	\;\textstyle\sum_{j}\frac{d_{j}\left(d_{j}+1\right)}{d_E\left(d_E+1\right)}\big((\Delta S_{jA})^{2}-\Psi'(d_E+2)+\Psi'(d_{j}+2)\;+\;\left(\av{S_{jA}}+\Psi(d_E+2)-\Psi(d_{j}+2) \right)^{2}\big)
\label{eq:varianceE}\\[.5em]
 &\textstyle +	\sum_{i\neq j}\frac{d_{i}d_{j}}{d_E(d_E+1)}\big(\left(\av{S_{iA}}+\Psi(d_E+2)-\Psi(d_{i}+1)\right)\left(\av{S_{jA}}+\Psi(d_E+2)-\Psi(d_{j}+1) \right)-\Psi'(d_E+2)\big)\,.\nonumber
\label{eq:SaverageE}
\end{align}
\end{widetext}
See App.~\ref{App:B} for a detailed derivation. In the limit of large dimension of each sector, $d_j\gg 1$, the average entanglement entropy takes the form
\begin{equation}
\textstyle \av{S_A}_{\!{}_E}=\sum_j \av{p_j} \,\av{S_{jA}}-\sum_j \av{p_j}\log \av{p_j}+O(1/d_E)\,.
\end{equation}
In the same limit, the variance $\textstyle (\Delta S_A)^2_{\!{}_E}=\av{S_{A}{}^{2}}_{\!{}_E}-\av{S_{A}}_{\!{}_E}^{2}$ goes to zero and the ratio of the two quantities scales as
\begin{equation}
\frac{(\Delta S_A)_{\!{}_E}}{\av{S_A}_{\!{}_E}}\sim \frac{1}{\sqrt{d_E}}\,.
\end{equation}
Therefore the average entropy of a subsystem is also its typical entropy. Fig.~\ref{fig:Histogram} shows that the normal distribution with mean \eqref{eq:SavOplus} and variance $\textstyle (\Delta S_A)^2_{\!{}_E}$ characterizes well the support of the distribution of the entanglement entropy of random pure states of energy $E$.


\emph{Discussion and applications}.---We can now answer the question posed in the introduction. \emph{As, for large systems, the standard deviation $\Delta S_A$ is much smaller than the average, we conclude that the average entropy $\langle S_A\rangle$ is also the typical value of the entanglement entropy of a subsystem. The conclusion holds both in the presence of a trivial and of a nontrivial center.}

The result applies to all systems where a subalgebra with a center arises. A notable case is the one of a free isolated quantum system prepared in an energy eigenstate. While interaction between subsystems is necessary for thermalization \cite{Deutsch1991, Srednicki1994, rigol.2008,d2016quantum, gogolin2016equilibration,Deutsch2018,goldstein_lebowitz_06, popescu_short_06, tasaki_98,polkovnikov2011colloquium,Vidmar:2017uux,Vidmar:2018rqk,Hackl:2018tyl,Vidmar:2017abc}, an arbitrarily small interaction is sufficient to select a typical energy eigenstate. We illustrate the relevance of our result with two examples.

A system of $N$ noninteracting spins with magnetic moment $\mu$ in a magnetic field $\mathrm{B}$ has Hamiltonian $H=\sum_i \mu_0 \mathrm{B}\, \sigma^z_i$. The energy eigenspace $\H(E)$ has dimension $d_E=N!/((\frac{N+n}{2})!(\frac{N-n}{2})!)$ with $n=E/\mu_0 \mathrm{B}$. We consider a subsystem $A$ consisting of $N_A$ spins. The eigenspace $\H(E)$ has the structure \eqref{eq:Hoplus} where the direct sum is over eigenvalues $\varepsilon$ of the Hamiltonian $H_A$ of the subsystem $A$. The average energy in $A$ is $\bar{\varepsilon}=\av{\Tr (H_A \rho_A)}_{\!{}_E}=E \,N_A/N$ which shows equipartition of the energy in a typical state. For $N\gg 1$, the average entanglement entropy of the subsystem \eqref{eq:SavOplus} evaluates to
\begin{equation}
\textstyle  \av{S_A}_{\!{}_E}\approx -\min(N_A,N_B) \sum_{i=\pm} \nu_i\log \nu_i\,,
\end{equation}
where $\nu_\pm=\frac{1}{2}(1\pm\frac{E}{N\mu_0 \mathrm{B}})$. This is the Page curve of the system and its variance is exponentially small in $N$. We observe that the entanglement entropy vanishes for the ground state $E=-N\mu\mathrm{B}$, increases with the energy up to $\min(N_A,N_B)\log 2$ in the middle of the spectrum, and then decreases indicating that the system behaves as if it had a temperature that is positive for low energy eigenstates and negative for high energy eigenstates \cite{Ramsey}. In fact for a small subsystem  ($N_A\ll N$) we can define a temperature as the variation of the entanglement entropy with respect to the typical energy
\begin{equation}
\textstyle \mathrm{kT}=\big(\partial \av{S_A}_{\!{}_E}/\partial \bar{\varepsilon}\big)^{-1}_{\!N,N_A}\approx\frac{-\mu_0 \mathrm{B}}{\mathrm{arctanh}\frac{E}{N\mu_0 \mathrm{B}}}\,,
\end{equation}
with $\mathrm{k}$ the Boltzmann constant. This is the familiar relation between the temperature and the energy in a paramagnetic system \cite{Balian}.
Fig.~\eqref{fig:pagecurve} shows the Page curve and its variance for a system of $N=10$ spins. The temperature of a small subsystem can be read from the slope of the Page curve.

 As a second example, we consider a free quantum field: the quantum electromagnetic field in a cubic box of volume $L^3$. Previous analysis have focused on the geometric entanglement entropy of a region of space  \cite{Sorkin:2014kta,Bombelli:1986rw, srednicki_93} and its renormalization \cite{Holzhey:1994we,Almheiri:2013wka,Bianchi:2014qua,Bianchi:2014bma}. Here we identify a different subalgebra of observables that better characterizes what can be measured \cite{Bianchi:2019pvv}. The Hilbert space of the quantum field $\H=\bigotimes_\lambda \H_\lambda$ is the tensor product of Hilbert spaces of discrete wavelength $\lambda=(2L/k_x,2L/k_y,2L/k_z)$ with $k_x,k_y,k_z\in \mathbb{N}$. A measuring device, such as an \emph{antenna} of length $\ell$, defines a subsystem $A$ corresponding to the discrete wavelengths $\lambda_A=(2\ell/k_x,2\ell/k_y,2\ell/k_z)$. If $\ell/L\leq 1$ is a rational number, the wavelengths $\lambda_A$ are a subset of the wavelengths $\lambda$. The Hilbert space $\H(E)$ of eigenstates with energy $E$ inherits the structure \eqref{eq:Hoplus}, where $\varepsilon$ is the energy of the antenna $A$.
When the box is large, i.e. when $EL\gg \hbar c$, we can compute the dimension $d_E$ of $\H(E)$ from the density of states and the occupation numbers of photons. We find $d_E\approx \exp\big(\frac{4\sqrt{\pi}}{3(15)^{1/4}}(\frac{E L}{\hbar c})^{3/4}\big)$. Analogously, the dimension of $\H_A(\varepsilon)$ for $\varepsilon \ell\gg \hbar c$ is $d_A(\varepsilon)\approx \exp\big(\frac{4\sqrt{\pi}}{3(15)^{1/4}} (\frac{\varepsilon\, \ell}{\hbar c})^{3/4}\big)$. 

The average energy in the subsystem $A$ is $\bar{\varepsilon}=\sum_\varepsilon \varepsilon \,\av{p(\varepsilon)} \approx E \ell^3/L^3$, which shows that in a typical state the energy of the subsystem is extensive and $E/L^3$ is the energy per unit volume. The average entanglement entropy of a subsystem with $\ell^3\leq L^3/2$ is
\begin{equation}
\av{S_A}\approx \log d_{A}(\bar{\varepsilon})\approx \textstyle\frac{4\sqrt{\pi}}{3(15)^{1/4}}\big(\frac{\bar{\varepsilon} \,\ell}{\hbar c}\big)^{\!3/4}\,.
\label{eq:Sblackbody}
\end{equation}
The result follows from  \eqref{eq:SavOplus} together with the fact that $\av{p(\varepsilon)}$ is sharply peaked at $\bar{\varepsilon}$ and therefore the  Shannon entropy contribution is negligible and the dominating term is the average entanglement entropy $\av{S_{\bar{\varepsilon}A}}$. As the variance is exponentially small in $EL/\hbar c$, this is also the typical value of the entropy in an energy eigenstate.

For $\ell\ll L$ we can also define a temperature from entanglement: 
\begin{equation}
\textstyle \mathrm{kT}=\big(\partial \av{S_A}_{\!{}_E}/\partial \bar{\varepsilon}\big)^{-1}_{\!L,\ell}\;\approx \big(\frac{15}{\pi^2} \frac{(\hbar c)^3\,E}{L^3}\big)^{\!1/4} \, .
\end{equation}
We note that this temperature does not depend on the subsystem size. 
In terms of this temperature, the entanglement entropy assumes the form of the familiar extensive formula for the canonical entropy of black body radiation. On the other hand, for large subsystems, the typical entanglement entropy is smaller than the thermal entropy and follows a Page curve qualitatively similar to the one in Fig.~\ref{fig:pagecurve},
\begin{equation}
\textstyle \av{S_A} \approx\frac{4\pi^2}{45}\big(\frac{\mathrm{kT}}{\hbar c}\big)^{\!3}\min(\ell^3,L^3-\ell^3)\,.
\end{equation}
This entropy is arising from the entanglement between the modes that the antenna can measure, i.e. the wavelengths $\lambda_A$, and the modes that it cannot measure. The unmeasured modes include both longer wavelengths and wavelengths shorter than $\ell$ that the antenna cannot couple to.

Our results show that, in small noninteracting systems prepared in a typical energy eigenstate, thermal properties can arise from entanglement. Recent experimental developments on measurements of thermalization in small isolated quantum systems, such as ultracold atoms in optical lattices~\cite{kinoshita2006quantum, islam_ma_15, kaufman_tai_16, lev2018}, might provide access  to the deviations from statistical mechanics predicted by the exact formulas for the average entropy \eqref{eq:SavOplus} and its variance \eqref{eq:varianceE} at fixed energy.


\PRLsep

\medskip

\emph{Acknowledgments}.---
The authors thank Neev Khera for discussions on the convergence of the $\delta \rho$ expansions,  Abhay Ashtekar for discussions on the structure of direct-sum decompositions, Pierpaolo Vivo and Lu Wei for correspondence. E.B. is supported by the NSF Grant No. PHY-1806428. P.D. is supported by the NSF Grants No. PHY-1505411, No. PHY-1806356 and the Eberly research funds of Penn State. 


\onecolumngrid

\appendix


\section{Detailed computation of the moments of the distribution $P(S_A)\dd S_A $}\label{App:A}
\label{sec:A}
Given a Hilbert space $\H$ of dimension $d$ and an orthonormal base $\ket{n}$ of $\H$, a normalized state $\ket{\psi}\in\H$ can be decomposed as $\ket{\psi}= \sum_{n=1}^d \psi_n \, \ket{n}$ with $\psi\in\mathbb{C}^d$ and $|\psi|=\sum_{n=1}^d |\psi_n|^2=1$. The uniform measure on $\H$ is then the measure on the unit sphere on $\mathbb{C}^d$,
\begin{equation}
\label{eq:uniform}
\dd \mu (\psi) = \frac{1}{Z} \delta\Big(1-\sum_{n=1}^d\left| \psi_n \right|^2 \Big)\prod_{n=1}^d \dd \psi_n \dd \overline\psi_n  \,,
\end{equation}
where $Z$ is a normalization constant. The average over states in $\H$ of functions of the reduced density matrix of a subsystem $A$ \cite{Page:1993df} is computed more efficiently using the measure over the eigenvalues of the density matrix induced by $\dd \mu (\psi)$, first computed by Lloyd and Pagels in \cite{Lloyd:1988cn}:
\begin{equation}
\label{eq:eigenvmeasure}
\dd \mu (\psi) \quad\longrightarrow\quad  \dd \mu \left(\l_{1},\ldots,\l_{d_{A}}\right)\;=\;
\tilde{Z} \, \delta\Big(1-\sum_{i=1}^{d_{A}}\l_{i}\Big)\Delta^2\left(\l_1,\ldots,\l_{d_A}\right) \prod_{k=1}^{d_{A}}\l_{k}^{d_{B}-d_{A}} \dd \l_k\,, 
\end{equation}
where $\Delta\left(\l_1,\ldots,\l_{d_A}\right)=\prod_{1\leq a< b \leq d_A} \left( \l_b - \l_a\right)$ is the Vandermonde determinant and $\tilde{Z}$ a normalization constant.

To compute the average of the entanglement entropy $S_A$ over the states in $\H$ we start by computing the integral
\begin{equation}
\label{eq:rhotor}
\av{\Tr(\rho_{A}^{r})}= \int \Tr(\rho_{A}^{r})\, \dd \mu (\psi) =\int\Big(\sum_{a=1}^{d_{A}}\l_{a}^{r}\Big)\dd \mu\left(\l_{1},\ldots,\l_{d_{A}}\right) \, .
\end{equation}
It is convenient to multiply \eqref{eq:rhotor} by a factor of $1$ written using the definition of the gamma function $\Gamma$,
\begin{equation}
\frac{1}{\Gamma\left(d_{A}d_{B}+r\right)}\int_{0}^{\infty}\zeta^{d_{A}d_{B}+r-1}e^{-\zeta}\dd\zeta=1 \, ,
\end{equation}
and perform a change of variables $q_a = \zeta \lambda_a$. Integrating over $\zeta$, we eliminate the delta function in the measure. The integration is over $d_A$ copies of the real positive line, $q_a\in \left[0,\infty\right]$, 
\begin{align}
\label{eq:rhoIntegral}
\av{\Tr\rho_{A}^{r}}&=
\frac{\tilde{Z}}{\Gamma\left(d_{A}d_{B}+r\right)}\int\Big(\sum_{a=1}^{d_{A}}q_{a}^{r}\Big)\Delta^2\left(q_1,\ldots,q_{d_A}\right)\prod_{k=1}^{d_{A}}q_{k}^{d_{B}-d_{A}}e^{-q_{k}}\dd q_{k}\,.
\end{align}
Each term of the sum over $a$ gives an equal contribution to $\av{\Tr\rho_{A}^{r}}$. Therefore, we write
\begin{align}
\av{\Tr\rho_{A}^{r}}&=\frac{\tilde{Z}\,d_A}{\Gamma\left(d_{A}d_{B}+r\right)}\int q_{1}^{r}\Delta^2\left(q_1,\ldots,q_{d_A}\right)\prod_{k=1}^{d_{A}}q_{k}^{d_{B}-d_{A}}e^{-q_{k}}\dd q_{k} \, .
\end{align}
The Vandermonde determinant can be computed using any monic polynomial. As noticed in \cite{Sen:1996ph,Dyer:2014}, it is convenient to compute the Vandermonde determinant using the generalized Laguerre polynomials $L_i^{(d_{B}-d_{A})}\left(q_a\right)$ since they form a complete and orthogonal basis of $L^{2}\left(\mathbb{R}^{+},q^{d_B-d_A}e^{-q}\,\dd q\right)$. We note that our convention for the Laguerre polynomials includes a factor $\left(-1\right)^{k}k!$, 
\begin{equation}
L_{k}^{(d_{B}-d_{A})}\left(q\right)=\left(-1\right)^{k}k!\sum_{r=0}^{k}\left(\begin{array}{c}
k+d_{B}-d_{A}\\
r+d_{B}-d_{A}
\end{array}\right)\frac{\left(-1\right)^{r}q^{r}}{r!}\, .
\end{equation}
Writing the determinant in terms of the fully antisymmetric tensor $\epsilon$ with $d_A$ indices $i_k=0,\ldots,d_A-1$, the integrals factorize
\begin{align}
\av{\Tr\rho_{A}^{r}}&=\tilde{Z}\sum_{i_k j_k}  \frac{d_{A}\epsilon_{i_{1}i_{2}\cdots i_{d_{A}}}\epsilon_{j_{1}j_{2}\cdots j_{d_{A}}}}{\Gamma\left(d_{A}d_{B}+r\right)}\int q_{1}^{r}\prod_{k=1}^{d_{A}}L_{i_{k}}^{(d_{B}-d_{A})}\left(q_k\right)L_{j_{k}}^{(d_{B}-d_{A})}\left(q_k\right)q_{k}^{d_{B}-d_{A}}e^{-q_{k}}\dd q_{k} \, .
\end{align}
All the integrals over $q_k$ with $k>1$ can be immediately computed using the orthogonality relation of the generalized Laguerre polynomials
\begin{equation}
\int_0^\infty L_{i}^{(d_{B}-d_{A})}\left(q\right)L_{j}^{(d_{B}-d_{A})}\left(q\right)q^{d_{B}-d_{A}}e^{-q}\dd q\;\propto\; \delta_{ij} \, .
\end{equation}
The two $\epsilon$ tensors are then contracted on all indices but the first one. Using the relation
\begin{equation}
\sum_{i_2,\ldots,i_{d_A}}\epsilon_{i_{1}i_{2}\cdots i_{d_{A}}}\epsilon_{j_{1}i_{2}\cdots i_{d_{A}}}\propto \delta_{i_1 j_1}\,,  
\end{equation}
and introducing a new normalization constant $\tilde{\tilde{Z}}$ to take care of the normalization of the measure \eqref{eq:eigenvmeasure} and of other proportionality factors, we write $\av{\Tr\rho_{A}^{r}}$ as
\begin{align}
\label{eq:rhoX}
\av{\Tr\rho_{A}^{r}}&= \frac{\tilde{\tilde{Z}}}{\Gamma\left(d_{A}d_{B}+r\right)}\sum_{ij}\delta_{ij} X_{ij} \left(r\right) \, ,
\end{align}
where we denoted $X_{ij}$ with $i,j=0,\ldots,d_A-1$ the integral
\begin{equation}
\label{eq:Xintegrale}
X_{ij}\left(r\right)=\frac{1}{\Gamma\left(i+1\right)\Gamma\left(d_B-d_A+i+1\right)}\int_{0}^{\infty}q^{r} L_{i}^{d_{B}-d_{A}}\left(q\right)L_{j}^{d_{B}-d_{A}}\left(q\right)q^{d_{B}-d_{A}}e^{-q}\dd q \, .
\end{equation}
To compute \eqref{eq:rhoX} we only need the values of $X_{ij}\left(r\right)$ with $i=j$. However, the computation of all matrix elements $i\neq j$ is useful for computing higher order moments and therefore is also computed here. The special case $X_{ij}\left(0\right)=\delta_{ij}$ is easy to obtain as it reduces to the orthogonality relation for the Laguerre polynomials. This is sufficient to determine the normalization constant, since
\begin{align}
d_A=\av{\Tr\rho_{A}^{0}}&= \frac{\tilde{\tilde{Z}}}{\Gamma\left(d_{A}d_{B}\right)}\sum_{ij}\delta_{ij} X_{ij} \left(0\right) = d_A \frac{\tilde{\tilde{Z}}}{\Gamma\left(d_{A}d_{B}\right)} 
\end{align}
requires $\tilde{\tilde{Z}}= \Gamma\left(d_A d_B\right)$. We then compute the integrals $X_{ij}\left(r\right)$ \eqref{eq:Xintegrale} for any $r\geq 0$ which appear in 
\begin{align}
\av{\Tr\rho_{A}^{r}}&= \frac{\Gamma\left(d_{A}d_{B}\right)}{\Gamma\left(d_{A}d_{B}+r\right)}\sum_{ij}\delta_{ij} X_{ij} \left(r\right) \, .
\end{align}
To keep the notation compact we denote $X(r)$ the $d_A \times d_A$ matrix with entries $X_{ij}\left(r\right)$. Using the generating function for the generalized Laguerre polynomials 
\begin{equation}
F\left(t,q\right)=\sum_{k}^{\infty}\left(-1\right)^{k}\frac{t^{k}}{k!}L_{k}^{(d_{B}-d_{A})}(q)=\frac{1}{\left(1-t\right){}^{d_{B}-d_{A}+1}}e^{-\frac{tq}{1-t}} \quad \text{and} \quad L_{i}^{(d_{B}-d_{A})}(q) = (-1)^i \left[\frac{d^{i}}{dt^{i}} F\left(t,q\right)\right]_{t=0}  \, ,
\end{equation}
we write \eqref{eq:Xintegrale} as derivatives respect to the parameters $x$ and $y$ of the integral of two generating functions:
\begin{align}
\label{eq:Xderivate}
X_{ij}\left(r\right)&=\frac{\left(-1\right)^{i+j}}{\Gamma\left(i+1\right)\Gamma\left(d_{B}-d_{A}+i+1\right)}\left[\frac{d^{i}}{dx^{i}}\frac{d^{j}}{dy^{j}}\int F\left(x,q\right) F\left(y,q\right) q^{d_B-d_A+r}e^{-q}\dd q \right]_{\substack{x=0,\\y=0\,}}\\
	&=\frac{\left(-1\right)^{i+j}\Gamma\left(d_{B}-d_{A}+r+1\right)}{\Gamma\left(i+1\right)\Gamma\left(d_{B}-d_{A}+i+1\right)}\left[\frac{d^{i}}{dx^{i}}\frac{d^{j}}{dy^{j}}\left(1-xy\right)^{-d_{B}+d_{A}-r-1}\left(1-x\right)^{r}\left(1-y\right)^{r}\right]_{\substack{x=0,\\y=0\,}}\, .
\end{align}	
We compute the derivatives explicitly by applying successively the Leibniz rule, obtaining a closed form for the integrals $X_{ij}\left(r\right)$
\begin{equation}
X_{ij}\left(r\right)=\frac{\Gamma\left(j+1\right)\Gamma\left(r+1\right)^{2}}{\Gamma\left(d_B-d_A+i+1\right)}\sum_{p=0}^{\;d_A-1}\frac{\Gamma\left(d_B-d_A+r+1+p\right)}{\Gamma\left(i-p+1\right) \Gamma\left(r+p-i+1\right)\Gamma\left(j-p+1\right)\Gamma\left(r-j+p+1\right)\Gamma\left(p+1\right)} \, .
\end{equation}
We conclude the computation for $\av{S_A}$ noticing that 
\begin{equation}
\av{S_A}=-\av{\Tr \rho_A \log \rho_A}= -\lim_{r\to 1}\partial_r \av{\Tr\rho_{A}^{r}}= -\lim_{r\to 1} \partial_r \frac{\Gamma\left(d_{A}d_{B}\right)}{\Gamma\left(d_{A}d_{B}+r\right)} \Tr X\left(r\right) \, .
\end{equation}
Taking the derivative and the limit is a long but straightforward calculation that can be done with the help of Wolfram's Mathematica. The result is the celebrated Page formula,
\begin{equation}
\av{S_A}=\Psi(d_{A}d_{B}+1)-\Psi(d_{B}+1)-\frac{d_{A}-1}{2d_{B}}  \, .
\label{eq:Sapp}
\end{equation}
The calculation of $\av{S_A{}^2}$ can be done following a similar strategy. We first compute
\begin{equation}
\av{\Tr(\rho_{A}^{r_1})\,\Tr(\rho_{A}^{r_2})}=\int\left(\sum_{a=1}^{d_{A}}\l_{a}^{r_1}\right)\left(\sum_{b=1}^{d_{A}}\l_{b}^{r_2}\right)\dd \mu\left(\l_{1},\ldots,\l_{d_{A}}\right)
\end{equation}
with $r_1>0$ and $r_2>0$. We perform a similar change of variables $q_a = \zeta \lambda_a$ and eliminate the delta function in the measure by integrating over $\zeta$. We obtain
\begin{align}
\label{eq:rhoIntegral2}
\av{\Tr(\rho_{A}^{r_1})\Tr(\rho_{A}^{r_2})}&=
\tilde{Z}\frac{1}{\Gamma\left(d_{A}d_{B}+r_1 + r_2\right)}\int\left(\sum_{a=1}^{d_{A}}q_{a}^{r_1}\right)\left(\sum_{b=1}^{d_{A}}q_{b}^{r_2}\right)\Delta^2\left(q_1,\ldots,q_{d_A}\right)\prod_{k=1}^{d_{A}}q_{k}^{d_{B}-d_{A}}e^{-q_{k}}\dd q_{k}\,.
\end{align}
The two sums can be expanded into two terms that can be integrated separately,
\begin{equation}
\label{eq:somme}
\left(\sum_{a=1}^{d_A}q_a^{r_1}\right)\left(\sum_{b=1}^{d_A}q_b^{r_2}\right)= \sum_{a=1}^{d_A}q_a^{r_1+r_2} +  \sum_{a\neq b=1}^{d_A}q_a^{r_1}q_b^{r_2} \, .
\end{equation}
The integral of the first term is completely analogous to the computation we just performed resulting in
\begin{align}
\frac{\tilde{\tilde{Z}}_1}{\Gamma\left(d_{A}d_{B}+r_1+r_2\right)}\sum_{ij}\delta_{ij} X_{ij} \left(r_1+r_2\right) \, ,
\end{align}
with a proportionality constant $\tilde{\tilde{Z}}_1$ to be determined later. The integral of the second term in \eqref{eq:somme} reduces to 
\begin{align}
\tilde{Z}\frac{d_A(d_A-1)}{\Gamma\left(d_{A}d_{B}+r\right)}\int q_{1}^{r_1}q_{2}^{r_2}\Delta^2\left(q_1,\ldots,q_{d_A}\right)\prod_{k=1}^{d_{A}}q_{k}^{d_{B}-d_{A}}e^{-q_{k}}\dd q_{k} \, .
\end{align}
Once again, we compute the Vandermonde determinant using Laguerre polynomials. This time the two fully antisymmetric tensors are contracted on all but two indices: 
\begin{equation}
\sum_{i_3,\ldots,i_{d_A}}\epsilon_{i_{1}i_{2}i_{3}\cdots i_{d_{A}}}\epsilon_{j_{1}j_{2}i_3\cdots i_{d_{A}}}\;\propto\; \delta_{i_1 j_1} \delta_{i_2 j_2} -\delta_{i_1 j_2} \delta_{i_2 j_1}  \, .
\end{equation}
We recast the integrals in terms of $X_{ij}\left(r\right)$ to obtain
\begin{align}
\frac{\tilde{\tilde{Z}}_2}{\Gamma\left(d_{A}d_{B}+r_1+r_2\right)}\sum_{ijkl}\left(\delta_{ij}\delta_{kl} - \delta_{il}\delta_{jk} \right)X_{ij} \left(r_1\right) X_{kl} \left(r_2\right)\, .
\end{align}
Summing the two contribution together we find
\begin{equation}
\av{\Tr\rho_{A}^{r_{1}}\Tr\rho_{A}^{r_{2}}}=\frac{1}{\Gamma\left(d_A d_B+r_{1}+r_{2}\right)}\left(
\tilde{\tilde{Z}}_1\Tr X\left(r_{1}+r_{2}\right)+
\tilde{\tilde{Z}}_2\Tr X\left(r_{1}\right)\Tr X\left(r_{2}\right)-
\tilde{\tilde{Z}}_2\Tr \left(X\left(r_{1}\right)X\left(r_{2}\right)\right)\right) \, .
\end{equation}
We fix the proportionality constants requiring $\av{\Tr\rho_{A}^{r}\Tr\rho_{A}^{0}}=d_A\av{\Tr\rho_{A}^{r}}$. The final expression is
\begin{equation}
\av{\Tr\rho_{A}^{r_{1}}\Tr\rho_{A}^{r_{2}}}=\frac{\Gamma\left(d_A d_B\right)}{\Gamma\left(d_A d_B+r_{1}+r_{2}\right)}\left(\Tr X\left(r_{1}+r_{2}\right)+\Tr X\left(r_{1}\right)\Tr X\left(r_{2}\right)-\Tr \left(X\left(r_{1}\right)X\left(r_{2}\right)\right)\right) \, .
\end{equation}
We conclude the computation for $\av{S_A^2}$ noticing that 
\begin{equation}
\av{S_A^2}=\av{\left(-\Tr \rho_A \log \rho_A\right)^2}= \lim_{\substack{r_1\to 1\\ r_2\to 1}}\partial_{r_1}\partial_{r_2}\av{\Tr\rho_{A}^{r_{1}}\Tr\rho_{A}^{r_{2}}} \, .
\end{equation}
We take the derivatives and the limits with the help of Wolfram's Mathematica. The variance of the entanglement entropy of a subsystem is defined as $(\Delta S_A)^2 = \av{{S_A}^2} - \av{S_A}{}^2$. Substituting our expressions for $\av{{S_A}^2}$ and $\av{S_A}$ we obtain the result reported in the main text paper:
\begin{equation}
(\Delta S_A)^2=-\Psi'\left(d_{A}d_{B}+1\right)+\frac{d_{A}+d_{B}}{d_{A}d_{B}+1}\Psi'\left(d_{B}+1\right)-\frac{\left(d_{A}-1\right)\left(d_{A}+2d_{B}-1\right)}{4d_{B}^{2}\left(d_{A}d_{B}+1\right)} \, .
\label{eq:DSapp}
\end{equation}

We can compute higher order moments of the entanglement entropy distribution employing the same strategy used to compute average and variance. As an example, we also report the computation for the third momentum $\mm_3=\av{\big(S_{A}-\av{S_{A}}\big)^{3}\,}$. First, we compute 
\begin{align*}
\av{\Tr\rho_{A}^{r_{1}}\Tr\rho_{A}^{r_{2}}\Tr\rho_{A}^{r_{3}}}=&\frac{\Gamma\left(d_{A}d_{B}\right)}{\Gamma\left(d_{A}d_{B}+r_{1}+r_{2}+r_{3}\right)}\Big(
\Tr X\left(r_{1}+r_{2}+r_{3}\right)+ 
 \Tr X\left(r_{1}+r_{2}\right)\Tr X\left(r_{3}\right)-\Tr \left(X\left(r_{1}+r_{2}\right)X\left(r_{3}\right)\right)\\
&\Tr X\left(r_{1}+r_{3}\right)\Tr X\left(r_{2}\right)-\Tr \left(X\left(r_{1}+r_{3}\right)X\left(r_{2}\right)\right)
+\Tr X\left(r_{3}+r_{2}\right)\Tr X\left(r_{1}\right)-\Tr \left(X\left(r_{3}+r_{2}\right)X\left(r_{1}\right)\right)\\
&\Tr X\left(r_{1}\right)    \Tr X\left(r_{2}\right)    \Tr X\left(r_{3}\right)+
 \Tr \left(X\left(r_{1}\right)X\left(r_{3}\right)X\left(r_{2}\right)\right)
+\Tr \left(X\left(r_{1}\right)X\left(r_{2}\right)X\left(r_{3}\right)\right)\\
&-\Tr X\left(r_{2}\right)\Tr \left(X\left(r_{1}\right)X\left(r_{3}\right)\right)
 -\Tr X\left(r_{1}\right)\Tr \left(X\left(r_{2}\right)X\left(r_{3}\right)\right)
 -\Tr X\left(r_{3}\right)\Tr \left(X\left(r_{1}\right)X\left(r_{2}\right)\right)\Big)
\end{align*}
Second, we take the limit of the derivatives respect $r_1$, $r_2$, and $r_3$
\begin{equation}
\av{S_A^3}=-\lim_{\substack{r_1\to 1\\ r_2\to 1\\ r_3\to 1}}\partial_{r_1}\partial_{r_2}\partial_{r_3}\av{\Tr\rho_{A}^{r_{1}}\Tr\rho_{A}^{r_{2}}\Tr\rho_{A}^{r_{3}}} \, .
\end{equation}
The computation is an herculean task but with the help of Wolfram's Mathematica we are able to simplify the exact formula for $\mm_3$:
\begin{align}
\mm_3\;=\;&	\Psi''\left(d_{A}d_{B}+1\right)-\frac{d_{A}^{2}+3d_{A}d_{B}+d_{B}^{2}+1}{\left(d_{A}d_{B}+1\right)\left(d_{A}d_{B}+2\right)}\Psi''\left(d_{B}+1\right)+\frac{\left(d_{A}^{2}-1\right)\left(d_{A}d_{B}-3d_{B}^{2}+1\right)}{d_{B}\left(d_{A}d_{B}+1\right)^{2}\left(d_{A}d_{B}+2\right)}\Psi'\left(d_{B}+1\right)
\label{eq:mu3app}\\[.5em]
&-\frac{\left(d_{A}-1\right)\left(2d_{A}^{3}d_{B}+3d_{A}^{2}d_{B}^{2}-4d_{A}^{2}d_{B}+2d_{A}^{2}+4d_{A}d_{B}^{3}-3d_{A}d_{B}^{2}+8d_{A}d_{B}-4d_{A}+10d_{B}^{2}-6d_{B}+2\right)}{4d_{B}^{3}\left(d_{A}d_{B}+1\right)^{2}\left(d_{A}d_{B}+2\right)}\,.\nonumber
\end{align}
When the subsystem $B$ is large, the result reduces to: 
\begin{equation}
\mm_3\approx -\frac{d_A^2-1}{d_A^3 d_B^3} \qquad \textrm{for}\quad d_B\gg 1\,.
\end{equation}
The skewness of a probability distribution is a measure of the asymmetry of the distribution and is defined as the ratio
\begin{equation}
\mm_3/\sigma^3  \approx   -\frac{\sqrt{8}}{\sqrt{d_A^2-1}} \, .
\end{equation}
The negative sign indicates a tilt of the distribution on the right of the median. 
 
\medskip

We give an example for a small system consisting of two qubits. The exact formulas \eqref{eq:Sapp}, \eqref{eq:DSapp}  at $d_A=2$, $d_B=2$ evaluate to
\begin{align}
&\textstyle \av{S_A}=\frac{1}{3}\simeq 0.333\,,\quad  \Smax=\log 2\simeq 0.693\,,\\[.5em]
&\textstyle \Delta S_A=\frac{1}{6}\sqrt{13-\frac{6}{5}\pi^2}\simeq 0.179\,,
\label{eq:numeric2}
\end{align}
which match the result of the numerical evaluation of the average over random pure states. 
 
 
\section{Uniform measure over a direct sum of Hilbert spaces}\label{App:B}
\label{sec:B}
We consider the eigenspace $\H(E)\subset \H$ with fixed energy $E$. $\H(E)$ has the structure of a direct sum of tensor products
\begin{equation}
\textstyle \H(E)=\bigoplus_{j=1}^J\Big(\H_A(\varepsilon_j)\otimes\H_B(E-\varepsilon_j)\Big)\,,
\end{equation}
where $\H_A(\varepsilon_j)$ and $\H_B(\varepsilon_k)$ are eigenspaces of given energy for the subsystems $A$ and $B$.

Energy eigenspaces of the subsystem $A$ are denoted $\H_{A}(\varepsilon_j)$ and have dimension $d_{jA}=\dim \H_{A}(\varepsilon_j)$. Similarly for subsystem $B$. The energy eigenspaces of the system have then the direct sum structure 
\begin{equation}
\textstyle \H(E)=\bigoplus_j \H_j(E)
\end{equation}
where the sector $\H_j(E)=\H_A(\varepsilon_j)\otimes \H_B(E-\varepsilon_j)$ has definite energy in each subsystem. We denote $d_j=\dim \H_j(E)$ the dimension of each sector, with $d_j=d_{jA}\, d_{jB}$ and $d_E=\sum_j d_j$ the dimension of $\H(E)$.

Any state $\ket{\psi, E} \in \H(E)$ can be written as $\ket{\psi,E} = \sum_j \sqrt{p_j} \ket{\phi_j}$ with $\ket{\phi_j}$ normalized to 1 and $\sum_j p_j = 1$. The coefficient $p_j$ can be interpreted as the probability of finding the state $\ket{\psi}$ in the sub-Hilbert space $\H_j(E)$. Without any loss of generality we can assume that a basis $\ket{n,E}$ of $\H(E)$ is adapted to the decomposition in $\H_j(E)$ meaning that $\ket{1,E},\ldots \ket{d_1,E}$ is a basis of $\H_1(E)$, $\ket{d_1+1,E},\ldots \ket{d_1+d_2,E}$ is a basis of $\H_2(E)$ and so on. Focusing for clarity on $j=1$, it is easy to see that
\begin{equation}
\dd \psi_1 \dd\overline{\psi}_1 \cdots \dd \psi_{d_1} \dd\overline{\psi}_{d_1}=  p_1^{d_1-1} \dd p_1\, \delta(\left| \phi_1 \right|-1) \dd\phi_{1,1}\dd\overline{\phi}_{1,1}\, \dd\phi_{1,2}\dd\overline{\phi}_{1,2}\,\cdots \dd\phi_{1,d_1}\dd\overline{\phi}_{1,d_1}\, =  p_1^{d_1-1}  \dd p_1\, \dd \mu (\phi_1 )\,.
\end{equation}
Repeating this decomposition on all the subspaces $\H_j(E)$, the uniform measure on $\H(E)$ can be written as the probability distribution of finding $\ket{\psi,E}$ in $\H_j(E)$ times the product of the uniform measures on $\H_j(E)$:
\begin{equation}
\textstyle \dd \mu (\psi ) = \dd \nu(p) \prod_j \dd \mu (\phi_j ) \ ,
\end{equation}
where 
\begin{equation}
\label{eq:distp}
\dd \nu(p) = \frac{1}{\mathcal{Z}} \delta\Big(1-\sum_{j=1} p_j  \Big)  \prod_j^J p_j^{d_j-1}\dd p_j \ .
\end{equation}
The normalization constant $\mathcal{Z}$ can be computed using a procedure similar to the one used in \eqref{eq:eigenvmeasure},
\begin{equation}
\mathcal{Z} = \int \delta\Big(1-\sum_{j=1} p_j  \Big)  p_j^{d_j-1} \prod_j \dd p_j = 
\frac{\prod_j \Gamma\left(d_j\right) }{\Gamma\left(d_E\right)} .
\end{equation}
The average and the variance of $p_j$ can be easily shown to be given by 
\begin{equation}
\label{eq:pj}
\av{p_j} = \frac{d_j}{d_E} \, , \qquad  (\Delta p_j)^2 =  \frac{d_j \left(d_E - d_j \right)}{d_E^2\left(d_E +1\right)} \, .
\end{equation}
The average of the Shannon entropy for the probability distribution \eqref{eq:distp} is given by
\begin{equation}
\av{S(p)}=-\av{\sum_j p_j \log p_j} = -\left[ \partial_r \av{\sum_j p_j^r}\right]_{r=1} = \Psi\left( d_E +1\right)  - \sum_j \frac{d_j}{d_E} \Psi\left( d_j +1\right)\, .
\end{equation}

\medskip

We illustrate the result with two examples. The first one consists in taking all equal dimensions  $d_j=d_E/J$. We find:
\begin{align}
&\av{p_j} = \frac{1}{J} \, , \qquad  (\Delta p_j)^2 =  \frac{J-1}{J^2 } \frac{1}{d_E+1} \, ,\\[1em]
&\av{S(p)}=\Psi\left( d_E +1\right)  - \Psi\left( d_E/J +1\right) \, \approx \,\log(J)\qquad \mathrm{for} \quad J\gg 1.
\end{align}

\medskip

As a second example, we consider the case where the dimension $d_J$ is much larger than the sum of all the others $d_J \gg d_R = \sum_{i=1}^{J-1}d_i=d_E-d_J$. In this case the exact formulas reduce to
\begin{align}
&\av{p_J} \approx 1- \frac{d_R}{d_J} \, , \qquad  (\Delta p_J)^2 \approx  \frac{d_R}{d_J^2} \\
&\av{p_i} \approx \frac{d_i}{d_J}  \, ,  \qquad   \qquad (\Delta p_i)^2 \approx  \frac{d_i}{d_J^2}\, \quad \text{ for } i\neq J\,,\\
&\av{S(p)}\approx \frac{1}{d_J}\Big(d_R+d_R\log d_J + \sum_{i=1}^{J-1} d_i \Psi\left( d_i +1\right)\Big) \, .
\end{align}
Note in particular that the average Shannon entropy goes to zero as $d_J\to \infty$. Furthermore if all the dimensions are large $d_j\gg 1$, we note that the average Shannon entropy equals the Shannon entropy of the average probability,
\begin{equation}
\av{S(p)}\approx \sum_{j=1}^{J} \av{p_j} \log \av{p_j} \, .
\end{equation}
The computation of $\Delta S(p)^2$ is straightforward but its expression is convoluted. We report here its leading order for $d_j\gg 1$, 
\begin{equation}
\Delta S(p)^2\approx \frac{1}{d_E}\sum_{j}\frac{d_{j}}{d_E}\left(1+\log\frac{d_j}{d_E}\right)^{2}-\frac{1}{d_E}\sum_{ij}\frac{d_{i}d_{j}}{d^{2}_E}\left(1+\log\frac{d_{i}}{d_E}\right)\left(1+\log\frac{d_{j}}{d_E}\right) \, .
\end{equation}
The variance of the Shannon entropy vanishes as $1/d_E$ for $d_j\to \infty$.

\twocolumngrid

\bibliographystyle{JHEP} 

\providecommand{\href}[2]{#2}\begingroup\raggedright\endgroup

\end{document}